\documentclass[fleqn]{llncs}
\usepackage{ist}
\usepackage{version}

\title{About Instruction Sequence Testing}
\author{J.A. Bergstra}
\institute{Informatics Institute, Faculty of Science,
           University of Amsterdam, \\
           Science Park~904, 1098~XH Amsterdam, the Netherlands \\
           \email{J.A.Bergstra@uva.nl}}

\begin{document}

\maketitle

\begin{abstract}
Software testing is presented as a so-called theme within which different authors and groups have defined different subjects
each of these subjects having a different focus on testing. A uniform concept of software testing is non-existent and the space of possible coherent perspectives on software testing, each fitting within the theme, is viewed as being spanned by five dimensions, each dimension representing two opposite views with a variety of intermediate views in between.

Instruction sequences are used as a simple theoretical conceptualization of computer programs. 
A theory of instruction sequence testing may serve as a model for a theory of software testing. Instruction sequences testing is considered a new topic for which definitions may be freely contemplated without being restricted by existing views on software testing. 

The problem of developing a theory of instruction sequence testing is posed. A survey is given of motivations and scenarios for developing a theory of instruction sequence testing. 
\end{abstract}

\section{Introduction}
\label{sect-introduction}

King \cite{King1976} states that program testing and program verification are two extreme
alternatives. He places symbolic execution in between of these two extreme alternatives. According to King testing involves presenting the program with inputs so that a run will
produce corresponding outputs. King explains that ``in testing a small sample of data 
that the program is expected to handle program is presented to the program. If the program is
judged to produce correct results for the sample it is assumed to be correct.''

I take from King's description of testing that it depends on experimental processes in which a real 
computing device plays a central role. In Singh \cite{Singh2012} one finds the following quite classic definition of program testing:
``Testing is the process of executing a program with the intent of finding faults''. 

Both views lead to immediate questions: concerning King's explanation I mention the following questions: (i) can a program produce any results,
(ii) is the program executed on a machine in order to produce these results, (iii) would manual execution
also qualify for testing, (iv) must it be exactly the program under test that is executed, or may it be a modified version, and how distant may that modification be from its original, (v) is there a need to 
contemplate an agent who performs the tests, (vi) do the intentions of such an agent matter?

The definition used in Singh's  book raises at least the following questions: 
(i)  is a test only successful if a fault is found,
(ii) is testing done to reveal faults, errors, or failures, (iii) can a test be performed manually, (iv) is execution needed or is interpretation sufficient to qualify as  a test, (iv) is it meant that the intent is to find faults in the program that is executed rather than in a modified program, (v) if no faults are found, and an inference is needed to infer the belief
that the program is correct, is making that inference a part of testing (in spite of it not being a part of 
executing), and if not what is the name of that reasoning phase, (vi) is it really necessary to involve something so remote as an intention and to presuppose the existence and activity of agents or perhaps even of non-artificial agents when 
defining program testing?

In spite of the temporal distance between the writing of \cite{King1976} and \cite{Singh2012}  one may hypothesize that King and Singh agree to some extent about the nature of
program testing. Middelburg \cite{Middelburg2010b} surveys the literature on program testing under the
assumption that it involves an experimental setup with a computing device. He draws two conclusions: (i) there
is a complete lack of theoretical papers that explain the experimental nature of testing (for instance what constitutes the difference between a test of a program and the use of a program), and (ii) there is no shred
of evidence why such experimental methods would be more effective or efficient, or in any sense better than 
so-called analytic methods such as verification and model-checking. 
I refer to \cite{Middelburg2010b} for original references to program testing and software testing.

A different view on program testing is found in Vouk \cite{Vouk2011} who states that ``(T)esting is the process of executing or evaluating a system, or a system component, by
manual or automated means to verify that it satisfies specified requirements or to
identify differences between expected and actual results or behaviors.'' In this definition all program analysis and verification methods are included, and the use of computers for experimentation is merely optional. I conclude that Vouk identifies program testing with software quality control. That view transpires also from \cite{Jangra2011} and from many other papers on program testing.

Yet another perspective on program testing, nowadays always called software testing, is that it constitutes a coherent
industrial activity, with specialized companies and well-trained staff. For instance, the UK based BCS specialist group on software testing (SIGIST) counts some 2500 members in 2011 and it has its own electronic newsletter, named ``The Tester'', with short but reasonably technical contributions such as for instance \cite{Martin2011}. 
Looking through 10 years of issues from The Tester one gets the impression that for BCS SIGIST testing is mainly about finding faults in software products and about obtaining confidence in the absence of such faults. This limitation is not a dogmatic matter, however,  and formal verification techniques would easily find their way to the BCS SIGIST community if that were considered effective because BCS SIGIST may at any time redefine its mission in terms of software quality assurance if that is considered more appealing. This view is consistent with not having a fixed meaning for ``program testing'' and to allow that phrase to refer to software industry's most favorite method of program quality engineering whatever techniques that approach may consist of.%
\footnote{I will not subscribe to this view, if only because assigning program testing this kind of dynamic meaning will make it very difficult to explain how to read the literature on testing. But the view has been promoted to a research level in work like \cite{RiunguTS2010} where a volume of practical experiences on Software Testing as a Service (STaaS) in the cloud is collected in order to build a so-called grounded theory following an approach suggested in \cite{Seaman1999}.}

In \cite{Hoare2010} Hoare states that testing and verification need not be considered opposite views, 
and that both profit from research based on the same logical methods, both can be used hand in hand an that it is not a researcher's task to consult a practitioner on when to use which of these methods. I conclude from these remarks that Hoare considers testing and verification to be different, but that he insists that a prolonged ideological debate about relative merits is futile. This
view takes its technical form in work like \cite{BeckmanNRSTT2010} and \cite{ChenTZ2011}. In the latter paper testing provides positive information from which an underapproximation of program behavior is obtained, that is a subset of the program semantics viewed as a relation, whereas verification is classified under methods for obtaining abstractions, an abstraction constituting an overapproximation, or superset in relational terms, of program behavior.

A way to understand program testing may be to contemplate the well-known slogan that ``program testing may reveal faults 
in a program but it may not reveal their absence''.  This is a remarkable position by all means. If one assumes that testing is an experimental process taking a physical form of a program as an input all measurement has a degree of uncertainty and for that reason leads to a probabilistic conclusion. In probabilistic terms, however, presence and absence of faults are functionally related, and the slogan becomes hard to defend. It follows that in order to appreciate the slogan one has to think in terms of certain knowledge. But then it is far from obvious that a single test execution can indeed provide the certainty that a fault is present. There are two difficulties: (i) it may the case that other system components than the program under test are faulty, and (ii) only if one has a mathematically precise theory predicting program behavior available, it can be predicted with certainty that some modification of a polinseq will  remedy the mismatch between the test outcome and its expectation. That mathematical/logical theory is more often than not unavailable to a tester.

\section{Program testing: which issues to tackle?}
In this paper my aim is twofold: to contribute to the theory of program testing, and to do so from the perspective of instruction sequences as a conceptual model of programs. That program testing still allows for at least two remarkable open issues has been established in \cite{Middelburg2010b} as mentioned above. And this state of affairs justifies further work in my opinion. 

\subsection{Scope of program testing}
The starting position that I have chosen for this paper includes the following statements: (i) the definition of program testing is still an open issue, and (ii) it is not clear to what extent it matters to have such a definition, so that (iii) it is unclear whether or not eventually an unambiguous and complete definition of program testing will emerge. A difficulty with these observations is that most authors on program testing don't share them. Indeed many papers on program testing contain some definition of program testing. The problem that I perceive lies in the fact that most authors active in the testing theme pay no attention to the fact that other authors in that same theme make use of quite different definitions. Most authors seem quite reluctant to appreciate the striking fact that precisely because their favored definition deviates from other author's preferred views on program testing some form of comparative justification is required for their own choice. The importance of program testing has been repeatedly stated on many places, though in the absence of a clear definition of program testing it is not so clear what software engineering activities can be counted as testing. Again this calls for further investigation: what is it that is generally considered to be so important? And what must be taught if one is committed to teaching software testing (see for instance \cite{AstigararraDLPW2010})?

\subsection{Towards solid underlying concepts}
Assuming that a theory of program testing needs to be based on well-defined concepts, the question arises to what extent this has been achieved in the theme of testing already. As it turns out additional clarification is needed for terms that are used in the description of testing. Here are some examples of that sort of matter:

\begin{itemize}
\item Many papers claim that program testing consists of exercising a program with a variety of inputs (e.g. \cite{CandeaBZ2010,MachadoVM2010}). This term exercising is not widely used outside testing so what does it mean? 
\item A related issue is the abundant usage of program execution as a description of some form of physical experiment. In view of \cite{Bergstra2011c} there is no basis for the assumption that instruction sequence testing must be based on instruction sequence execution.%
\footnote{The instruction sequence testing activity is best viewed as a structured matter with its own particular workflow. In IEEE Standard 829--1998 for Software Test Documentation a specific workflow for the testing activity is presupposed. I refer to \cite{MachadoVM2010} for a discussion of the mentioned workflow. The testing activity at large may involve ``executing a test'' as one of its phases, which itself may consist of putting an instruction sequence into effect rather than executing it.}
The strictly more liberal notion of ``putting instruction sequences into effect''  suffices for explaining instruction sequence testing. In \cite{Calame2008} testing is based on the even more liberal 
notion of simulation, whereas in \cite{BrandanBriones2007} testing is seen as experimenting with a system in an attempt to find errors.  It is claimed in \cite{Watkins2009} that unit testing was done manually and before coding in a waterfall-model based software process.
\item A remarkable unclarity concerns the notion of test, which according to some authors (e.g. \cite{WeyukerOstrand1980}) coincides with a set of inputs each of which can be used for ``executing a test''. Other authors would rather speak of a test set in this case. 
\item  It is unclear whether program testers are looking for failures, errors (as in \cite{BarbosaSCM2008}) or faults (as in \cite{Singh2012}). Only failures are ever observed when testing a program, and errors that are not failures may be observed by either testing a program in a modified environment or by testing a modified program. 
\item To what extent the reasoning from observed dynamic errors to hypothesized faults in an instruction sequence is a  part of program testing requires further investigation. It seems plausible to have in included in testing. But that requires clarity about its mechanics.%
\footnote{This form of so-called causal reasoning is a topic of ongoing research, see for instance \cite{ChenTZ2011}. There is no common logical analysis of this form of reasoning, which is surprising because it must be done whenever a test is to be interpreted.}
\end{itemize}

\subsection{Theory of program testing}
The question what constitutes a theory of program testing is worth further reflection, in spite of a literature on testing comprising many thousands of papers. Typical occurrences of a theory of program testing are found in \cite{WeyukerOstrand1980,GG75a,Gou83a} and in \cite{StaatsWH2011}. What these papers do is to analyze formal connections between a spectrum of terms, at the same time ignoring the physical background of the use of computers involved. In spite of the fact that these theories are useful and have had and still have significant impact, their strength in terms of capability to generate or  to accommodate support definitions of all terms involved is  limited.

\subsection{Objectives of program testing}
The final question which I want to raise about instruction sequence testing, assuming that one thinks in terms of fault analysis by putting the instruction sequence or a slightly modified version of it into effect is the following: should one conceive of testing as an experimental investigation potentially leading to conclusions which  must be considered more important than the conclusions which are drawn from theories made for predicting system behavior. In other words, if a postcondition $\phi$ has been proven for an instruction sequence $x$, and after a test run $\neg \phi$ is observed, does this observation refute the proof system if the proof is logically valid, or the proof if the proof system is beyond doubt. Here one find the simplest interpretation of instruction sequence testing in philosophical terms: an empirical process which is essential for validating or refuting the soundness of proof methods for dynamic properties of instruction sequences. Remarkably all proof systems come with the implicit assumption that empirical soundness can be taken for granted on logical grounds. In the world of quantum computing such a view is hard to imagine, however.

If one understands a proof system for program correctness as a theory capable of predicting system behavior in some cases, the need for testing emerges at once: validating or refuting this theory. Looking at it this way, testing is always and in principle in the lead, while program proving is no more than systematic inductive inference on the basis of testing outcomes. Is this view reasonable, and if so, how has it happened that testing became less regarded than verification in principled circles?

\subsection{Revision of program testing definition: an implausible task}
I will refer to program testing as a theme which by definition incorporates all serious writing and practice performed under the heading of ``program testing'' and of ``software testing'', irrespective of underlying differences of opinion concerning the meaning of these phrases. An author of a specific work in the theme of program testing will treat it as a subject (also called program testing), which implies that the author has some meaning of the phrase ``program testing'' in mind, a meaning which may or may not have been made explicit by the author, either in full text and detail 
or by means of references to previous works.

I consider the state of affairs that so many authors within the theme of program testing simply ignore the foundations of the subject assumed by other authors working in the same theme both disappointing and remarkable. Indeed often authors are very clear about the importance of the subject as they conceive it and at the same time they prove in writing that the theme is definitely not sufficiently important to merit even a minimal and perhaps critical representation of other schools of thought in the same theme. Is seems to be the case that the theme fragments into different subjects which are fighting each other primarily by ignoring another's existence.

Given this state of affairs I take any objective of sharpening the definition of program testing to be premature at this stage. At the same time I find all scholarly writing, including text books for academic courses,  on the theme of program testing that emphasizes the importance of the theme and at the same time simply focuses on a private definition of program testing, without making mention of other views, as fundamentally defective. In my view the fraction of program testing literature that shows signs of this kind of defect is high, which strikes me as an unattractive state of affairs.%
\footnote{In \cite{MaesDeVries08} Maes and deVries explain the methodological aspects of work that aims at questioning or changing a so-called dominant interpretation. It seems to be a dominant interpretation that authors on program testing need not spend effort on detailing the basic concepts of their work in view of the directions take by other authors. I would be happy to change that situation and in view of \cite{MaesDeVries08} theoretical work rather than empirical work is the more plausible way to go to that end.}

Two approaches to work on testing may contribute to a change of this state of affairs: (i) working towards better definitions of program, including comparative studies, which I will do on the neutral playground of ``instruction sequence testing'', and (ii) criticizing works on program testing which feature the mentioned defect and analyzing in each case the extent to which the defect is detrimental for the credibility and for the effectiveness of the work.%
\footnote{The work proposed can be classified as design work concerning definitions and theory. As a proposed design task it deviates from what Avital et. al. in \cite{AvitalLBBD2006} have proposed under the name ``design with a positive lens''. This may be a minor defect from some perspective, but I agree that it would be quite useful if ``foundational'' work on program testing can be wrapped in the cloths of design with a positive lens. In related work in \cite{AvitalTeni08} simulation is given a central role for obtaining so-called generativity in design. Generativity refers to a ``deep''  and self-reinforcing match between design activity and the underlying motivational context.}

\section{Instruction sequence testing: a neutral territory}
In order to make the paper independent of judgements about the state of affairs in program testing I will work towards a theory of instruction sequence testing. I hold that I am free to contemplate different possibilities for defining instruction sequence testing, even in the case that the meaning of program testing and of software testing would be fixed already. Thus, assuming that instruction sequences are a simple conceptual model for programs a theory of instruction sequence testing may be a model for a theory of program testing, and also for a theory of software testing assuming one wishes to distinguish program testing from software testing. The objective of this paper is not to provide a definition of instruction sequence testing, because at this stage I simply don't know if  that is a feasible goal. 
Rather the objective is to map out the context in which this definition may be designed if at all.

Looking at the matter from a different angle: I will consider ``instruction sequence testing theory''  to be a new and neutral theme where definitions may be provided without any automatic impact on the definition of program testing. Whenever I write "instruction sequence testing'' that should be read in the context of ``instruction sequence testing theory''. This convention helps to avoid, at least in principle, that the plausible assumption that every instruction sequence is a program  introduces a logical connection between instruction sequence testing and program testing, thereby prematurely restricting the freedom to experiment with definitions of instruction sequence testing. If this is still vague ``instruction sequence theory'' is supposed to be understood as in e.g. \cite{BL02a,BM07e,BP04a,BM08,BZ08,Bergstra2011c}.

\subsection{Inseq, polinseq, and polinsequt}
In order to have a more ``realistic'' setting I will work with polyadic instruction sequences, where a polyadic instruction sequences comprises a number of instruction sequences in a single entity of which the behavior is defined and potentially used. Polyadic instruction sequencing has been introduced in the setting of instruction sequences in \cite{BM08}, whereas at the semantic level of threads the same kind of modularity has been developed under the name of poly-threading in \cite{BM11a}.
I will abbreviate instruction sequence to inseq and polyadic instruction sequence  to polinseq. Systems based on control via instruction sequences will in the majority of cases make use of polyadic instruction sequences, so that ``inseq testing'' must generally be understood as ``polinseq testing''.  Similarly ``program testing''
stands for ``program family'' testing. Software testing need not be adapted in this way because software is closed under ``composing families of''.

I will refer to a polinseq that is being tested as a target polinseq or a polinseq under test, for which the abbreviation polinsequt
is proposed.

\subsection{Five dimensions of polinseq testing as a theme}
I will distinguish five aspects, or dimensions, where two extreme views on polinseq testing serve as the bounds of a range of intermediate views. This give rise to a five dimensional space of possible but focusse interpretations of polinseq testing as a subject, and by taking the union of all possibilities it also describes a plausible extension of polinsec testing as a theme. Different subjects can be combined into more liberal ones by taking them together. Thus he powerset of this collection of possible but focussed subjects on testing constitutes a container of the possible (though less sharply focussed) subjects that can be conceived within the theme of software testing. This determines  TTS, the Testing Theory Space, as a container of possible extensions of a testing theory, from which a theory designer may try to make a selection. Each of these degrees of freedom has been taken from the literature on program testing, where similar variations can be observed.%
\footnote{In \cite{GelperinHetzel1988} the observation is made that different objectives of testing have been dominant in different temporal phases. According to that view at no time all of the possibilities of the spectrum below have been simultanously plausible, and testing fashion makes some kind of motion through TTS. I have the impression, however, that old habits don't easily disappear and that an accumulation of aspects of the mission of testing take place instead of a mere transformation.

In \cite{Beizer1990} a maturity level structure for the software testing process of an organization is proposed which is similar to the stages in the historic development of \cite{GelperinHetzel1988}. The remarkable consequence of this view is that conventional views on testing are labeled as immature, or at least as less effective. in any case, if one adopts Beizer's viewpoint then testing cannot be defined in terms of the tester's intentions. Because these intentions are the main discriminating factor for the different maturity levels. In \cite{AmmannOffutt2008} Beizer's views are said to be followed, which is said to be consistent with the following definition formulated in \cite{AmmannOffutt2008}: `` testing is evaluating software by observing its execution''. This definition leads to the questions: Why execution? Which software exactly is executed? What is the difference with use?}

\subsubsection{Focus on inference from controlled experiments versus focus on all means of analysis.}
This dimension concerns the range from (i) conceiving a test as a process of obtaining information (e.g. finding errors, faults and failures) about a polinseq (the polinsequt or polinseq under test) by putting the polinseq into effect with selected inputs (the test set, or test sample) as inputs, to (ii) application of all means of analyis, notably including formal verification, model checking, code walk-through, and usage scenario and risk analysis, for obtaining an assessment of polinseq quality. For instance so-called symbolic execution is a technique which fits in this range.%
\footnote{The well-known slogan that testing can reveal the presence of polinseq errors but cannot reveal their absence can only be understood if a focus on controlled experiments is assumed.}

\subsubsection{Focus on target polinseq quality (what is wrong) versus focus on polinseq production process quality 
(why it went wrong).}
This dimension contrasts testing as fact finding about a specific polinseq (the target polinseq or polinsequt), with fact finding about the underlying production process. 
Structural testing methods have a focus on the intrinsic rationale of a polinseq as an industrial product. Obviously if some instructions in a polinseq cannot be accessed during any run this fact merits attention. The fact that polinseqs with redundant parts are at all produced may be considered a weakness of the used polinseq production process, which in its turn may be used as incentive to improve that process.

\subsubsection{Focus on internal polinseq quality (how it works) versus focus on the quality of polinseq functionality (what it does).}
This dimension concerns the range between assessment of the rationale of the target polinseq (if this is the solution, how good is that solution) and assessment of the match between polinseq behavior and the underlying requirements, (is this indeed a solution) in the case that  any requirements are known.

\subsubsection{Focus on the target polinseq (product quality) versus focus on mutations of the target polinseq (testing  process quality).} This contrast accomodates the empirical fact that many tests consisting of putting a polinseq into effect over a sample of inputs in fact involve the putting into effect of ad hoc modifications of the target polinseq.

\subsubsection{Focus on validity (requirements quality) versus focus on correctness (implementation quality).} This range acknowledges that fact that validity and correctness are independent notions and that testing may be motivated by analyzing various combinations of these two aspects.

\subsection{Polinseq quality threats}
This paragraph  is a reworking and an extension of the introduction of \cite{Laski1997} using the wording of a Wikipedia page on system dependability. It provides information about the usage of several terms, but without providing definitions thereof. Of these so-called quality treats polinseq faults constitute the major enemy which testing is hoped to oppose, under the assumption that mistakes will always be made. The other treats provide a context for appreciating the notion of a fault.

Threats  specify way in which a polinseq driven system can be affected causing a drop in system dependability, in other words: threats reveal a defective polinseq quality. There are three main terms concerning the occurrence of threats that must be clearly understood:
\begin{description}
\item{\em Mistake.} A (polinseq production) mistake is a human action which causes the presence of a fault in a polinseq.%
\footnote{``Mistakes'' includes the occurrence of  failures of tools that the human polinseq-author is using.}
\item{\em Fault.} A fault in a polinseq (which often used to be referred to as a bug for historic reasons) is a defect of that polinseq. The presence of a fault in a polinseq may or may not lead to a failure when the polinseq is put into effect.%
\footnote{Understanding the notion of a fault seems to be a matter of intuition rather than of clear analysis and definition. It seems to be an underlying assumption that all non-artistic artifacts resulting from human engineering activities are fault-prone, and each type of artifact brings with it its own ideology of what constitutes a fault. Fault is a static notion, in contrast with error and failure which always depend on a specific run.}
\item{\em Error.} An error is a discrepancy between the intended behaviour of a polinseq, when being put into effect, and its actual behaviour inside the system boundary. Errors occur at runtime when the system enters an unexpected state due to the activation of a fault. 
\item{\em Failure.} A failure is an event when a system displays behaviour that is contrary to its specification. This behavior must be visible for an external observer. An error may not necessarily cause a failure, for instance an exception may be thrown by a system but this may be caught and handled using fault tolerance techniques so the externally observable operation of the system will conform to the specification.
\end{description}
It is important to note that failures are observed at the system boundary. They are merely errors that have propagated to the system boundary and have become observable.  In general a fault, when activated, can lead to an error (which is an invalid state) and the invalid state representing an error may lead to one or more successor errors and eventually to a failure, which is by definition an observable deviation from the specified behaviour at the system boundary. If an error is not followed by a failure the error is said to have been {\em masked}.

\subsubsection{Non-dependability related quality issues.}
Code compactness measures the size of a polinseq in relation to the functionality that it must represent. In order to quantify code compactness some software metric is needed, and in particular a polinseq metric is needed. The total number of instructions is an option for such a metric. In any case code compactness increases if functionality is preserved while lowering the measured metric. Now code compactness is almost unrelated to dependability, it is rather related to constructability: can one load specific functionality on a given device at all? Obtaining an assessment of code compactness by means of putting a polinseq into effect on a set of test inputs seems impossible, or at least implausible. What we find is that some quality characteristics of polinseqs are not mere reformulations of being bugfree.

\subsection{What is a polinseq fault?}
It constitutes a difficulty that no unambiguous technical definition of fault seems to be available.%
\footnote{It seems reasonable to consider an instruction sequence faulty whenever if dissatisfaction about it emerges from contemplating requirements that were available before its engineering and if in addition a strategy of enacting systematic change is preferred over a performing complete redesign, as a way to resolve the given dissatisfaction. A significant contribution to the definition of a program fault is given by Janusz Laski  in \cite{Laski1997}. That rather unique paper contains a definition of what it means for a software component to contain a fault, or rather to be faulty. The amazing fact about this definition is that it requires the presence of a formalized operational semantics as well as the presence of a proof system for correctness. If this paper were the best option for defining program incorrectness (viewed as the presence of a fault) then that leads to the remarkable hypothesis that both an axiomatic semantics and an operational semantics are conceptual prerequisites for a proper understanding of what a fault is, thus turning a program fault into a quite sophisticated notion.}

The presence of a fault appears if making a minor  local change, called a remedy, in the polinseq removes the errors that led to its identification without leading to any new errors. Judging the latter is by no means easy. Local changes to a polinseq may involve (i) the insertion of novel instruction to a polinseq, (ii) the removal of existing instructions from the polinseq, and (iii) the modification of existing instructions within the polinseq. Each of these steps are instruction level modifications.

Inferring a fault includes finding a remedy, because only in the presence of proposed remedy it can be assessed that a fault has been identified.%
\footnote{Some numerical bound to the number of instruction level changes permitted  for contructing a remedy needs to be given in advance. A $k$--remedy consists of a series of at most $k$ successive instruction level modifications. Then one may speak of $k$--remedy visible faults, that is faults admitting a $k$--remedy. With increasing $k$ this notion becomes increasingly meaningless of course.}

\subsubsection{Faults from an polinseq author's perspective.} Assuming that a polinseq is produced (written) by an author or by a team of authors, the simplest understanding of a fault is that when writing an inseq composing the polinseq the author does not carry out adequately what (s)he intends to achieve. Thus, when pointed out the subsequence of instructions which  contain the alleged fault the author would recognize the mismatch between polinseq engineer intentions and the polinseq engineering product at hand. These intentions concern a plan on how to implement and intended functionality by means of a polinseq.

\subsubsection{Faults from a polinseq reviewer's perspective.} It is possible that the polinseq engineer
is mistaken about how an implementation of a known functionality may be achieved. Then (s)he will be unable to 
detect the fault when pointed out its location, but an independent reviewer might see what is wrong. If a local modification exists that removes the consequences of this fault while not introducing any novel errors, the problem may still be considered a fault. If no local change can be found that remedies the error an overall design problem has been detected which can hardly be understood as a collection of faults.

\subsubsection{Faults as an error diagnosis.} Finally the observation of failures may lead to the detection of errors. Finding the cause of errors may lead to additional insights, essentially like the reviewer's insight obtained by inspecting a polinseq as a text. The diagnostic work following the detection of an error may or may not lead to the highlighting of a family of limited subsequences of some of the inseqs contained in the given polinseq as potential locations of a fault. 

\section{Polinseq quality engineering}
I take quality engineering to comprise all systematic activities that can contribute to achieving product and process quality. On the basis of this explanation, I conclude that polinseq testing falls entirely within polinseq quality engineering. Questionable is whether polinseq testing and polinseq quality enigeering can be identified. The five dimensional topic space on polinseq testing covers all of polinseq quality engineering. Thus, taking that space as a the only working hypothesis, this question cannot be resolved.  

A reasonable decomposition of polinseq quality engineering is obtained by taking the working methods as the most important criterion. That leads to the following three compartments of the theme of polinseq quality engineering. These parts concern products as well as production proceses.

\begin{description}
\item{\em Polinseq experimentation.} Experimental investigation of polinseqs is primarily based on making observations on running versions of a poliseq, including simulations in modified conditions and of modified polinseqs.
\item{\em Polinseq logic.} In polinseq logic or polinseq mathematics the assertions that one intends to find are formalized and 
the evidence sought  takes the strength of a mathematical proof. That is, one states and proves theorems about a polinseq
 (or specific modifications of it), and its behavior. This may involve informal mathematics, highly formalized mathematics, proof checking and model checking.
\item{\em Polinseq analysis.} Analysis includes: various forms of informal human comprehension, peer reviews, maturity level assessments, code walk-throughs, various metrics, comparisons with benchmark polinseq portfolios.
\end{description}

\subsection{Three indications of polinseq testing}
I will use a concept indication as a pointer to where (the definition of) a concept may be found. 
An indication lacks the clarity of a definition,  indications are not unique, and different indications may be of different proximity to the concept. Indications can be investigated before definitions are provided. Indications may serve as requirements for definitions.

My view on polinseq testing as a subject is that it constitutes  a strict and coherent subset of polinseq quality engineering. Polinseq testing makes essential use of experimental methods whereas experiments need not play role in other chapters of polinseq quality engineering. Here are three indications of polinseq testing. 

\begin{itemize}
\item 
Polinseq testing constitutes a part of polinseq quality engineering.
\item
Polinseq testing is strictly contained in polinseq quality engineering.
\item 
Polinseq testing is the experimental part of polinseq quality engineering.
\item 
Polinseq testing is the experimental part of polinseq quality engineering, that is, polinseq testing is polinseq experimentation, where the other parts of Polinseq testing is the experimental part of polinseq quality engineering are polinseq logic and polinseq analyisis.
\end{itemize}

I am inclined to subscribe to each of these indications of polinseq testing. Rather than designing further refinements of these indications I will now consider different ways to classify aspects of  polinseq quality engineering.

\subsection{Options for refining the indication of polinseq testing}
In principle all five dimensions for polinseq testing as a theme testing allow a degree of freedom for a definition of polinseq testing as well. This leads to an unhelpful explosion of options. By considering only a fraction of these options, structured in a comprehensible way the design space acquires a workable form, that will suffice for a preliminary phase of the definitional work on polinseq testing, which consists of refining the indication for polinseq testing obtained thus far.

I will structure the different scenarios in a tree, first making the distinction between product, process, and method (of reasoning and analysis). 
\subsubsection{Polinseq testing is about product quality.} In this case faults are always found in a product (i.e. polinseq).This scenario subdivides as follows:
\begin{description}
\item{\em Testing stands for quality engineering.} In this kind of definition testing is used for all tasks and tools for program quality engineering. The justification for talking about testing if something more general is actually meant 
lies in the dominant role of testing more narrowly conceived. In other words: executing a test is only a part of testing. And most, but not all testing methods involve executing a test on some polinseq.
\item{\em Testing stands for a limited collection of experimental methods.} Narrowly conceived testing is always about experimental techniques essentially  involving some form of, execution, interpretation, or simulation (in general: putting into effect). The following aspects of product quality are in focus of polinseq testing.
\begin{description}
\item{\em Judging the existence polinseq failures for a given polinseq.} Detection of failures leads to detection of errors and to detection of faults.
\item{\em Assessment of the presence  polinseq errors for a given polinseq.} This is the same activity as detecting failures though in a modified setting where system boundaries are made more narrow by reporting internal data outside during or after a polinseq run. Indeed every error becomes a failure in an appropriately expanded setting.
\item{\em Identification of polinseq faults on the basis of observed polinseq errors.} This is the central inference in controlled experiment based polinseq testing.
\item{\em Assigning a subjective probability to the absence of faults in a polinseq.} Testing results both in fault finding, which leads to repair and renewed testing, and to the testing process terminating without producing direct evidence for remaining faults in a polinseq. After testing has terminated there may well be statistical evidence for the existence of
faults which are yet undetected. For that reason the subjective probability of the existence of residual faults need not 
be zero after a prerelease testing phase has come to an end.
\end{description}

Here a further subdivision is meaningful.
\begin{itemize}
\item Testing a polinseq involves putting that polinseq into effect only. Here is a further ramification:
	\begin{itemize}
	\item Putting a polinseq  into effect for testing purposes  is done in the intended execution architecture.
	\item Putting a polinseq  into effect may be done within a range of dedicated execution environments, some of 		which are specifically designed for testing purposes.
	\end{itemize}
\item Testing a polinseq may involve putting into effect a range of modifications of the given polinseq.
\end{itemize}
\end{description}

\subsubsection{Polinseq testing is about production process quality.} Here the idea is that polinseq faults are
caused by an inadequate production process. Rather than repairing an individual polinseq fault the objective is to modify the production process so that future products don't feature similar faults. The following subdivision is possible:
\begin{description}
\item{\em High quality product generation.}
In this case the objective of testing is to address all product quality issues which can be influenced via adequate production processes.
\item{\em Bug free product generation.} The more limited objective is that a polinseq engineering process is designed in such a way that it minimizes the number of faults in each of its products. Testing is still aiming at finding faults but rather than updating the product when a fault has been spotted the process is revised in order to avoid repetition.
\end{description}

\subsubsection{Polinseq testing is about the validation of polinseq logic systems and  polinseq analysis methods.} In this case the prime contribution of experimental work on polinseq quality comprises the development and validation of reasoning methods for polinseq logics and of polinseq analysis methods for. More specifically:

\begin{description}
\item{\em Validation/refutation of proof systems.} Empirical work is required if a theory (which is formalized in terms of a polinseq logic)  for predicting system behavior must be evaluated.
\item{\em Heuristics for developing proof methods.} In a setting where or only few proof rules predicting system behavior have been found, experimentation helps to accumulate information from which useful additional proof rules may be guessed by way of hypothesis. This way testing comprises the ground work for a systematic and explorative investigation of system behavior.
\item{\em Investigation of boundary conditions.} Even the soundest proof system for instruction sequence correctness depends on boundary conditions concerning its execution environment (see \cite{BP04a}). These boundary conditions may not be expressible in the formal notation of a proof system so that experimental work is be 
needed to get an understanding of where the boundaries of validity of a prediction actually lie.
\item{\em Confirming limits of prediction.}
It can be imagined that some types of computing systems develop in such a way that on principled grounds, that is explained by a theory of relevant system dynamics perhaps comparable to
quantum mechanics  or to chaos theory, no prediction of certain kinds of behavior is possible. Validating the theoretical account of this claim of impossibility must ultimately involve experimental work.
\end{description}

\subsubsection{A fourth indication of polinseq testing.}
I subscribe to the idea that a refinement of the indication that {\em polinseq testing is the experimental part of polinseq quality engineering} is found by stating that in addition {\em it is primarily about product quality}, 
with obtaining consequences for production process
quality as a derived results only. 

Validation of polinseq logics and of polinseq analysis methods may require experimental work. That work, however, is to be captured under polinseq quality engineering research, which is not itself included in polinseq quality engineering. This separation of the validation of polinseq logic, as being a matter of research,  from the validation of a single polinseq as being a matter of practice, is not without problems. It seems to imply that polinseq testing cannot occur as a part of polinseq testing research. But  that implies that polinseq testing does not exist, unless polinseq engineering has grown into a significant practice, a development which is not foreseen.

A way out of this dilemma can be found in the following two steps: (i)  to capture the generation and validation of polinseq logics as well as the development of new polinseq analyis methods under fundamental polinseq engineering research, while the development of specific polinseqs, including the pertinent quality engineering is considered  a matter of applied polinseq engineering research. It then comes about that polinseq testing contributes to applied polinseq engineering research, but not to fundamental polinseq engineering research. Next, (ii) it must be understood that polinseq engineering research may have an applied part, even if there is no intended application for the polinseqs that arise during the research work. The justification for labeling a part of the research on polinseq engineering as applied is that its objective is to model known software engineering processes. It would be labeled fundamental research only if novel methods, having no clear predecessors in software engineering, are investigated.

\subsection{Potential objectives of polinseq testing theory}
A theory of polinseq testing will both provide and be dependent on a particular definition of testing. Different definitions give rise to different theories, which all may be included in the program testing theme. Below I will list some objectives that may be pursued when developing a theory of polinseq testing.
\begin{description}
\item{\em Attractive extension of polinseq theory.} A valid objective is to extend the theory of polinseqs with an attractive chapter on testing, which adds to the comprehensiveness of polinseq theory.
\item{\em Applications of testing to polinseq implementation efforts.} A second objective is to provide implementation mechanisms for instruction sequences and to extend these with theory based testing tools, thus enhancing what can be done with polinseqs.
\item{\em Transfer from polinseq testing theory to program testing theory.} It is conceivable (if not explicitly sought) that polinseq testing theory produces some conclusions which can be transformed to a more general setting of program testing in such a way that novel insights on program testing are found.
\item{\em Creation of new program testing tools and techniques.} If a transfer to program testing can be found, this transfer may demonstrate its value if its leads to new tools and techniques for known testing problems.
\end{description}
I consider the first objective in the listing above to be the most plausible and achievable one, the other objectives being only accessible and realistic after the first one has been confronted with some degree of success. It is conceivable that different definitions of polinseq testing can form the basis of different and equallly attractive theories of polinseq testing. That makes the choice of a definiton of polinseq testing rather important. 

To what extent a theory of polinseq testing can match the criteria on theory development as given in \cite{Sifakis2011} is an unclear matter. It is probably plausible and relevant (as meant in \cite{Sifakis2011}), but it may be deficient in terms of being well-defined according to the spirit of \cite{Sifakis2011}.

\section{Why working on instruction sequence testing?}
Apart from the fact that instruction sequence testing may be defined without 
constraints emerging from the tradition of different definitions for program testing,
thus creating a formidable, and necessary degree of freedom,
it is of course very helpful as well if working on the basis of instruction sequences will
give rise, after an appropriate transfer process, to tangible progress for program testing research as well. 
At the same time the approach can have disadvantages and that matter will be discussed first

\subsection{Potential disadvantages of the approach}
Because instruction sequence notations are quite low
level it may be the case that faults are spread over a set of instructions
in a less coherent fashion than would have been the case for programs written in a high level program notation. Thus.
assuming that inferring a fault from an observed failure or error is a major obstacle, 
that obstacle may be smaller in a high level program notation, which may be considered to constitute a disadvantage of instruction sequence based research on software testing.

Further, a program fault is often considered to be the result of a programming mistake. Now so-called high-level
program notations have been defined in such a way as to minimize the occurrence of programmer mistakes. This
can't be said for instruction sequence notations which have been designed for maximal clarity of the explanation of the
mechanics of putting them into effect in a variety of execution environments. Indeed instruction sequencing,
the equivalent of programming for instruction sequences, may not provide a productive setting for the investigation of author mistakes, because of an underestimation of human factors in notational design, at least in 
comparison with program notations that are in daily use in practice.

A third disadvantage of theorizing about software testing by means of the thought experiment of polinseq testing is as follows: testing software is a process which precedes taking the decision to use that software (for a definition of decision taking see \cite{Bergstra2011d}). At some stage test results are (expected to be)  such that these do not stand in the way of release or of use or of making a  decision to that extent. For polinseqs, for which no realistic examples exist, it may not be possible to create a sufficientlly realistic setting of decison taking to allow to take this aspect adequately into account.

\subsection{Conjectured advantages}
Before a the project to develop a theory of polinseq testing has actually been carried out, no decisive statement can be issued about the advantages and disadvantages of that project in comparison to working on the basis of one or more known program notations. For that reason the advantages mentioned below have a conjectural status by necessity.  Due to its basis in instruction sequence theory, development  of the theory of polinseq testing, will enjoy advantages of the following kind:
\begin{description}
\item{\em Flexible program notation.} Using instruction sequences based on program algebra
one may find a large spectrum of different but related polinseq notations.
\item{\em Flexible execution environment.} By separating the polinseq notation from its intended 
execution architecture it becomes possible to introduce a notation independent range of variation 
which allows a systematic investigation of the cause effect relation between faults and failures, 
unavailable when working with a conventional program notation.
\item{\em Clarity concerning ``execution''.} For polinseqs a theory of putting them into effect can 
be introduced which sheds light on the various forms of execution, exercising, simulation, 
interpretation, running, managed execution, which may be needed for an explanation of the 
experimental process (if any) that constitutes the body of the polinseq testing process.
\item{\em Precise definition of performance notions.} Performance notion such as the 
maximal internal delay
of a polinseq can be easily introduced. Testing of performance on a low level of 
abstraction can be studied in meticulous detail.
\item{\em Very clear operational semantics.} By combining projection semantics and operational semantics a very precise operational meaning can be assigned to each polinseq. This is very useful for defining and analyzing the cause effect relationship between faults and errors/failures. Restriction of the number of steps in a run to an upper bound can be easily defined and implemented. 
\item{\em Mathematical formalism.} Results about (pol)inseq testing can in principle be formulated
in a mathematical style, without making use of any known program notation. Such formulations may be stable
with respect to changes in fashion for programming methods and programming style.
\end{description}

\subsection{How to get started: still the chicken or the egg dilemma concerning definitions?}
The simplest way to start with the development of a theory of polinseq testing is to provide a definition of a polinseq test, if only a provisional one. The starting point for providing a provisional definition might be a bundle of slogans like: ``program testing is executing a program in order to find faults in it''. These may be transformed into a polinseq adapted form, like for instance 
``polyadic instruction sequence testing is executing a
polyadic instruction sequence in order to find faults in it''. At the same time one may initially dismiss slogans like: ``program testing covers the entire area of software quality''. Having thus collected a set of accepted polinseq adapted slogans and in addition to that a collection of dismissed polinseq adapted slogans, one may focus on making these or similar statements 
precise in a coherent framework of concepts with adequate definitions.

This seems quite unproblematic, but in fact it is rather problematic. The problem is this: rather than working towards a coherent conceptual framework with precise definitions one might directly develop an instruction sequencing toolbox and start to acquire experience with activities that are very likely to pass the test of being tests in a wide range of meanings that theorist  might want to assign to the notion of an instruction sequence test. I will use naive polinseq testing for a practice that emerges without backing from an conceptual and theoretical framework which has been developed already. One cannot maintain that only after having developed the theoretical concepts practical experience can be acquired. In other words, naive polinseq testing need not be a bad thing, on the contrary it may be a  good thing. This leads to the following questions:

Is it legitimate, is it possible, and is it advisable to work out definitions of polinseq testing without having gained experience with naive polinseq testing in advance?

I suggest the following answers:  yes, yes and no. On general ``philosophical'' grounds gaining hands-on experience cannot be considered a necessary prerequisite for theory development in any area.  But intuitively it seems advisable to gain experience in naive polinseq testing before working towards a definition of polinseq testing. Equipped with the mentioned experience one may provide a ``better'' (more grounded, more informative, richer) definition, whereas having gained this experience as a preparatory step is unlikely to do any harm in connection with that objective.  But  gaining experience in naive polinseq testing is not a trivial task and that very fact may lead someone to ignoring the advice just formulated, if (s)he is sufficiently sure on how to proceed in theory mode, or sufficiently keen to proceed in that direction in spite of the risk sprouting from a lack of hands on experience. 

\section{Conclusions}
\label{sect-conclusions}
A survey of the complications with the notion of program testing has been presented. It has been argued that working towards a ``better'' definition of program testing is unlikely to be a rewarding strategy.  Instead the suggestion is presented  to work out a theory of
instruction sequence testing (or more precisely polinseq testing). That theory may serve  as a model of a theory of program testing, which may lead to applications on program testing, and it may also provide a useful extension of the existing work on instruction sequences. Potential merits of this course of action have been outlined, as well as a risk-analysis concerning the value of a polinseq testing theory as a model of program testing theory.

\bibliographystyle{spmpsci}
\bibliography{TA}

\end{document}